\documentclass[11pt,twoside
]{article}

\usepackage{baaa2010}
\usepackage{graphicx}
\usepackage{psfrag}
\usepackage{amssymb}
\usepackage[spanish,activeacute,english]{babel}
\usepackage[latin1]{inputenc}
\usepackage[T1]{fontenc} % Computer Modern (CM) fonts
\usepackage{ae,aecompl} % and: dvips -Pcmz -Pamz macros_aaa.dvi
\usepackage{latexsym}
\usepackage{verbatim}
\usepackage{amsmath}
\usepackage{amsfonts}
\usepackage{amssymb}
\usepackage{wasysym}
\usepackage[colorlinks=true,dvips]{hyperref}
\begin{document}
%%%%%%%%%%%%%%%%%%%%%%%%%%%%%%%%%%%%%%%%%%%%%%%%%%%%%%%%%%%%%%%%%%%%%%%%%%
%%%% SELECCIONE EL IDIOMA EN QUE SE ESCRIBE EL ARTÍCULO:              %%%%
%\myselectspanish
\myselectenglish
%%%%%%%%%%%%%%%%%%%%%%%%%%%%%%%%%%%%%%%%%%%%%%%%%%%%%%%%%%%%%%%%%%%%%%%%%%
\vskip 1.0cm
%\markboth{ J.J. Clari\'a \& F.A. Bareilles }%

\pagestyle{myheadings}
%%%% DESCOMENTE LA LINEA QUE DESCRIBE EL CARACTER DE SU TRABAJO       %%%%
\vspace*{0.5cm}
%\noindent TRABAJO INVITADO 
\noindent PRESENTACIÓN ORAL
%\noindent PRESENTACIÓN MURAL
%\noindent RESUMEN 
\vskip 0.3cm
\title{The antenna DSA 3 and its potential use for Radio Astronomy}

%\title{ Template paper for publication in the Bulletin of the 
%Argentinian Astronomical Association with instructions for the use of 
%\LaTeX{}}

\author{P. Benaglia$^{1}$, 
N. Casco$^{1}$,
S. Cichowolski$^{2}$,
A. Cillis$^{2}$, 
B. Garc\'{\i}a$^{3}$,
D. Ravignani$^{3}$,
E. Reynoso$^{2}$, 
G. de la Vega$^{3}$}

\affil{%
  (1) Instituto Argentino de Radioastronom\'{\i}a (IAR)\\
  (2) Instituto de Astronom\'{\i}a y F\'{\i}sica del Espacio (IAFE)\\
  (3) Instituto de Tecnolog\'{\i}as en Detecci\'on y Astropart\'{\i}culas (ITeDA)\\
}

\begin{abstract} 
The European Space Agency (ESA) will inaugurate its third Deep 
Space Antenna (DSA 3) by the end of 2012. DSA 3 will be located 
in Argentina near the city of Malargüe in the Mendoza province. 
While the instrument will be primarily dedicated to communications 
with interplanetary missions, the characteristics of its antenna 
and receivers will also enable standalone leading scientific 
contributions, with a high scientific-technological return. 
We outline here scientific proposals for a radio astronomical 
use of DSA 3.
\end{abstract}

\begin{resumen}
La Agencia Espacial Europea (ESA) inaugurar\'a, a finales 
de 2012, la  tercera  antena de espacio profundo (DSA 3), 
en suelo argentino (Malarg\"ue, Mendoza). El instrumento se dedicar\'a 
principalmente a comunicaciones con misiones interplanetarias. 
% Sin embargo, hasta un 10% del tiempo total de observación está 
% garantizado por la ESA para la utilización de la antena por 
% nuestro país. 
Dadas las caracter\'{\i}sticas de la antena y receptores, 
con la DSA 3 se podr\'an realizar contribuciones científicas de 
punta, con un alto retorno cient\'{\i}fico-tecnol\'ogico. 
Aqu\'{\i}\ se delinean propuestas científicas para su uso 
radioastron\'omico.
\end{resumen}

\section{Introduction}

The purpose of DSA 3 is to provide support to ESA 
interplanetary missions, like Mars Express, Venus 
Express, Rosetta, and the upcoming BepiColombo. 
DSA 3 will have an antenna of 35 m in diameter ($D$) and 
will work receiving and sending 
radio signals in two frequency bands, $X$ and $K_{\rm a}$ (about 8 and 32 
GHz respectively). 
To make contact with missions typically located at three million kilometers and beyond, the use of low-noise 
amplifiers cooled to cryogenic temperatures is mandatory, along with 
highly accurate pointing and calibration. The facility will also have 
devices for tracking, telemetry modulation and demodulation, telecommand 
and data, radiometric and meteorological measurements.

ESA already has two of these stations: DSA 1 in New Norcia 
(Australia) since 2002 and DSA 2 in Cebreros (Spain) since 2005. 
These antennas, also of $D = 35$ m, are currently the largest 
ones operated by ESA. DSA 3 will complement 
ESA deep-space network, ensuring around-the-clock coverage for their 
interplanetary missions. 
In return for harboring DSA 3 within its territory, ESA offers Argentina the
use of up to 10\% of the observing time, which will represent an outstanding
advantage for the local scientific community.
Given the unique technical features of the instrument, 
this fraction of time can be utilized for first-level 
research in the radio astronomy field and in astrophysics in general.

\section{Proposals for scientific use of the Deep Space Antenna 3}

DSA 3 will observe at centimeter and millimeter wavelengths 
with 1 arcminute angular resolution at 32 GHz and 4.5 arcminutes at 8 GHz. 
Receivers cooled up to 20 K will be able to obtain accurately calibrated 
data with very low noise, at much shorter time scales than other 
instruments. Scientific proposals would be selected according to their 
academic excellence by peer reviewers. The host country 
has researchers formed in Radio Astronomy, that can make 
an intensive use of DSA 3 and, simultaneously, train new radio 
astronomers using the facility. Forefront research lines that 
can be carried out during the host country observing time are 
presented below.\\

\noindent {\bf Unidentified gamma-ray sources.} Last-generation gamma-ray 
telescopes, such as the FERMI satellite, and the H.E.S.S. I and II arrays
(Namibia), continuously monitor the southern sky and are detecting 
thousands of sources of very high energies (e.g., Fermi-LAT 2nd Source
Catalog 2011 and
Aharonian et al. 2008, respectively) with angular resolutions 
comparable to those 
of the DSA stations. An important part of the detected sources  ($\sim 
30$\%) cannot be identified as they do not have a counterpart in other energy 
ranges. The existence of a physical relationship between radio and 
gamma-ray emission has been known for decades (Ginzburg \& Syrovatskii 
1965). DSA 3 will be a critical tool to determine the nature of the 
unidentified sources, and to study the processes that contribute to the 
radiation. Just to underline the importance of the subject, we remark
the amount of resources allocated to instruments looking at 
high-energy sources, the number of scientists commited to solve the underlying 
problems and the amount of related articles published on a daily 
basis.\\

\noindent {\bf The radio galaxy Centaurus A.} Cen A harbors the closest 
supermassive black hole. However, its proximity implies a large source 
angular extension, which prevents us from observing it with radio 
interferometers at intermediate frequencies (10 - 30 GHz, Israel 1998, 
Israel et al. 2008, etc). An instrument like DSA 3 can obtain data from 
all Centaurus A in a few days of observation. For the sake of comparison, 
the Australia Telescope Compact Array took about 2 years to map Cen A at 
1.4 GHz (Feain et al. 2011). Observations at different frequencies 
could also provide information on the emission mechanisms. 
Moreover, the study of the magnetic fields  and the relativistic 
particles involved will be possible by the fully polarimetric 
information that DSA 3 will supply.\\

\noindent {\bf Variability of Active Galactic Nuclei.} 
With DSA 3 it will be possible to perform variability studies 
of brightness and polarization degree of Active Galactic 
Nuclei (AGN) emission in a very efficient way.
Studies on time scales of hours to years will report on the 
accretion process on the galaxy's central black hole, still poorly known. 
Multi-frequency data from a large sample of AGN will allow the study of the 
physical processes that support the variability of individual objects, 
and also of the differences between different classes of AGN (e.g., Hovatta 
et al. 2008, Gliozzi et al. 2009). The advantage of time series analysis 
is that the information is gathered independently of the AGN type, 
and that the results complement those found by means of spectroscopic studies.\\

\noindent {\bf Supernova remnants and HII regions.}
The measuremet of the fluxes both at the $X$ and $K_{\rm a}$ bands can be 
used to discriminate the nature of the source observed; in particular, 
SNRs and HII regions. Over 50\% of the Galactic SNRs are 
above the limit of detectability offered by the DSA 3, even at 32 GHz. 
For those of larger angular size, it will be possible to study spatial 
variations of the spectral index and to measure directly the direction of the 
magnetic field. Parameters of the HII regions will be obtained from the 
measured fluxes, such as the mass of ionized gas and the electron 
density. $K_{\rm a}$-band observations will be essential to elucidate, for 
example, why, although expected, there is no radio continuum 
emission towards a number of O-type stars, or 
how significant is the contribution of 
the surrounding background emission to the measured flux.\\

\noindent {\bf Physics and chemistry of proto-planetary clouds.} 
Interstellar dark clouds are the places where stars and planets are 
formed. The knowledge of their chemical composition is 
essential to understand how matter in the
Universe evolves to planets and life (e.g., Cernicharo et al. 2008). 
The study of these clouds helps not only to 
determine chemical abundances, but also their growth (e.g. Ohishi \& Kaifu 
1998). The instruments used to date, besides having a very high 
oversubscription, are all in the Northern Hemisphere 
(NRO Nobeyama-45m, 100-m Effeslberg, Onsala OSO-20m, IRAM-30m dishes), 
with a declination limit $\sim -25^0$. In addition, the 
observing bands are above 75 GHz for all but NRO. An instrument 
like DSA 3 will be ideal, and in many cases the only option, 
to study southern dark clouds, given its great 
sensitivity, adequate angular resolution and fast coverage.\\

\noindent {\bf ``Flickering'' and the interstellar medium.} The 
interstellar material present in the line of sight towards 
another galaxy, produces an effect of scintillation or ``flickering'' of 
the extragalactic radiation. The study of this phenomenon 
is an important tool to gain information about the small scale inhomogeneities
in the electronic component of the interstellar medium
(e.g., Bochkarev \& Ryabov 2000). The variability in brightness is produced by 
the refraction of light that traverses, at high speed, a hot gas 
permeated by shock waves. Studies of variability -in this case, 
extrinsic to the object- can be conducted in different time 
scales (days to years) with DSA 3.\\

\noindent {\bf Rotating radio transients.} With DSA 3 it will 
also be possible to 
search for transient events in radio waves just as with NASA deep 
space antennas (Buu et al. 2011). The so-called RRATs (Rotating 
RAdio Transients) have been recently discovered (McLoughlin et al. 2006, 
Lorimer et al. 2007) as sources of very intense pulses that may be of 
extragalactic origin. The physical description of these rare objects is 
proving to be a challenge to such an extent that they are considered 
testers of basic physics and astrophysics.

\section{Scientific and technological return}
                   
The excellent quality of the state of the art components of DSA 3 guarantees 
benefits to radio astronomy that are of great importance in leading 
areas. Since the bulk of the investment is provided by ESA, the use of 
DSA 3 for radio astronomy ensures a maximum benefit/return to science and 
technology with a minimum of cost/investment. Furthermore, 
observing during all the host country time will re-position 
Argentina as a world class player in the Radio Astronomy arena.

Besides the advancement of the knowledge frontier in the outlined 
lines of research, some of the additional benefits that can be 
attained through DSA 3 include:

\begin{itemize}

\item Insertion of Argentina in the radio astronomy research mainstream.

\item International collaborations with top-level groups.

\item Publication of results in international high-impact journals.

\item Training of technicians and engineers in the priority area of Information and Communications Technologies (ICTs).

\item Training of scientists and engineers at graduate and postgraduate level.

\item Technological developments in electronic engineering and software.

\item Transfer of technology in communications, instrumentation and control, digital signal processing, frequency and time metrology.

\end{itemize}

\acknowledgements
%agradecimientos
The authors are grateful to the ITeDA staff and to Nora Loiseau.
P.B. thanks support from  PICT 2007-00848, ANPCyT.

\begin{referencias}

\reference Aharonian et al. 2008, A\&A, 353
\reference Bochkarev \& Ryabov 2000, New Astron. Rev., 44, 375
\reference Buu et al. 2011, IEEE, 99, 5, 889
\reference Cernicharo et al. 2008, ApJ, L83
\reference Feain et al. 2011, ApJ, 740, 17
\reference Ginsburg \& Syrovatskii 1965, ARA\&A, 3, 297
\reference Gliozzi et al. 2009, ApJ, 703, 221
\reference Hovatta et al. 2008, A\&A, 485, 51
\reference Israel 1998, ARA\&A, 8, 237
\reference Israel et al. 2008, A\&A, 483, 741
\reference Lorimer et al. 2007, Science, 318, 777
\reference Mc Loughlin et al. 2006, Nature, 439, 817
\reference Ohishi \& Kaifu 1998, Faraday Discussions No. 109, 205
\reference The Fermi-LAT Collaboration 2011, arXiv:1108.1435

\end{referencias}

\end{document}